\documentclass[showpacs,preprintnumbers,amsmath,amssymb,preprint]{revtex4}
\usepackage{graphicx}
\usepackage{dcolumn}
\usepackage{bm}

\begin{document}

\title{Absence of the  Fifth  Force Problem in a Model with Spontaneously
 Broken Dilatation Symmetry}


\author
{E. I. Guendelman \thanks{guendel@bgu.ac.il} and A.  B. Kaganovich
\thanks{alexk@bgu.ac.il}}
\address{Physics Department, Ben Gurion University of the Negev, Beer
Sheva 84105, Israel}

\date{\today}

\begin{abstract}
A scale invariant model containing dilaton $\phi$ and dust (as a
model of matter) is studied where the shift symmetry
$\phi\rightarrow\phi +const.$ is spontaneously broken at the
classical level due to intrinsic features of the model. The
dilaton to matter coupling "constant" $f$ appears to be dependent
of the matter density. In normal conditions, i.e. when the matter
energy density is many orders of magnitude larger than the dilaton
contribution to the dark energy density, $f$ becomes less than the
ratio of the "mass of the vacuum" in the volume occupied by the
matter to the Planck mass. The model yields this kind of
"Archimedes law" without any especial (intended for this) choice
of the underlying action and without fine tuning of the
parameters. The model not only explains why all attempts to
discover a scalar force correction to Newtonian gravity were
unsuccessful so far but also predicts that in the near future
there is no chance to detect such corrections  in the astronomical
measurements as well as in the specially designed  fifth force
experiments on intermediate, short (like millimeter) and even
ultrashort (a few nanometer) ranges. This prediction is
alternative to predictions of other known models.


\end{abstract}

 \renewcommand{\baselinestretch}{1.6}


\maketitle

\section{Introduction}
\label{Introduction}

Possible coupling of the matter to a scalar field can be the
origin of a long range force if the mass of the scalar particles
is very small. It is well known since the appearance of the
Brans-Dicke model\cite{Brans-Dicke} that such "fifth" force could
affect the results of tests of General Relativity (GR). In more
general cases it may entail a violation of the Einstein's
equivalence principle.
 A possible existence of light scalar particles
interacting to matter could also give rise to testable
concequences in an intermediate or submillimeter or even  shorter
range depending on the scalar mass. Numerous, many years lasting,
specially designed experiments, see for example
\cite{Experiments-1}-\cite{Whas_Einstein_right}, have not revealed
so far any of possible manifestations of the fifth force. This
fact, on each stage of the sequence of experiments, is treated as
a new, stronger constraint on the parameters (like coupling
constant and mass) with hope that the next generation of
experiments will be able to discover a scalar force modifying the
Newtonian gravity. This is the essence of the fifth force problem
in the "narrow sense"\footnote{As is well known, other
implications of the light scalar  generically may be  for
cosmological variations of the vacuum expectation value of the
Higgs field, the fine structure constant and other gauge coupling
constants. However in this paper we study only the strength of the
 fifth force itself. }.

In this paper we demonstrate that it is quite possible that the
fifth force problem in such narrow sense does not exist. Namely we
will present a model where the strength of the dilaton to matter
coupling measured in experimental attempts to detect a correction
to the Newtonian gravity  turns out so small that at least near
future experiments will not be able to reveal it. On the other
hand, if the matter is very diluted then its coupling to the
dilaton may be not weak. But the latter is realized under
conditions not compatible with the design of the fifth force
experiments.

The idea of the existence of a light scalar coupled to matter has
a well known theoretical ground, for example in string
theory\cite{GreenSchwarzWitten} and in models with spontaneously
broken dilatation symmetry\cite{Peccei-1987},\cite{Peccei-1989}.
 The fifth force problem has acquired
a special actuality in the last decade when the
quintessence\cite{quint} and its different modifications, for
example coupled quintessence\cite{Amendola},
k-essence\cite{k-essence}, were recognized as successful models of
the dark energy\cite{DEreview}. If the amazing observational
fact\cite{coinc} that the dark energy density is about two times
bigger that the (dark) matter density in the present cosmological
epoch is not an accidental coincidence but rather is a
characteristic feature during long enough period of evolution,
then the explanation of this phenomenon  suggests that there is an
exchange of energy between dark matter and dark energy. A number
of models have been constructed with the aim to describe this
exchange, see for example\cite{Amendola},
\cite{Amendola-2}-\cite{Alimi}  and references therein. In the
context of scalar field models of the dark energy, the
availability of this energy exchange implies the existence of a
coupling of the scalar field to dark matter. Then immediately the
question arises why similar coupling to the visible matter is very
strongly suppressed according to the present astronomical
data\cite{Whas_Einstein_right}. Thus the resolution of the fifth
force problem in its modern treatment should apparently consist of
simultaneous explanations, on the ground of a fundamental theory,
of both the very strong suppression of the scalar field coupling
to the visible matter and the absence of similar suppression of
its coupling to the dark matter.

One of the interesting approaches to resolution of the fifth force
problem known since 1994 as "the least coupling principle" based
on the idea\cite{LeastCoupling} to use non-perturbative string
loop effects to explain why the massless dilaton may decouples
from matter. In fact it was shown that under certain assumptions
about the structure of the (unknown) dilaton coupling functions in
the low energy effective action resulting from taking into account
the full non-perturbative string loop expansion, the string
dilaton is cosmologically attracted toward values where its
effective coupling to matter disappears.

The astrophysical effects of the matter density dependence of the
dilaton to matter coupling   was studied in 1989 in the context of
a model with spontaneously broken dilatation symmetry in Ref.
\cite{Peccei-1989}. However in this model the effect is too weak
to be observed now. Another way to describe the influence of the
matter density on the fifth force is used in the Chameleon
model\cite{Chameleon} formulated in 2004. The key point here  is
the fact that the scalar field effective potential depends on the
local matter density $\rho_m$ if the direct coupling of the scalar
field to the metric tensor in the underlying Lagrangian is assumed
like in earlier models \cite{Damour-1},\cite{Veneziano}. Therefore
the position of the minimum of the effective potential and the
mass of small fluctuations turn out to be $\rho_m$-dependent.  In
space regions of "high' matter density such as on the Earth or in
other compact objects, the effective mass of the scalar field
becomes so big that the scalar field can penetrate only into a
thin superficial shell of the compact object. As a result of this,
it appears to be possible to realize a situation where in spite of
a choice for a scalar to matter  coupling of order unity, the
violation of the equivalence principle is exponentially
suppressed. However, for objects of lower density, the fifth force
may be detectable and the corresponding predictions are made.

One should note that the model of Ref. \cite{GK1} with the matter
density dependence of the effective dilaton to matter  coupling
was constructed in 2001 without any specific conjectures in the
underlying action intended to solve the fifth force
problem\footnote{The more detailed description of the model
\cite{GK1} and its results, including new effects as neutrino dark
energy which appear when the fermion density is very low, are
presented in Ref. \cite{GK2}.}. The resolution of the fifth force
problem appears as a result which reads: 1) The local effective
Yukawa coupling of the dilaton to fermions in normal laboratory
conditions equals practically zero automatically, without any fine
tuning of the parameters. The term normal laboratory conditions
means that the local fermion energy density is many orders of
magnitude larger than the dilaton contribution to the dark energy
density. 2) Under the same conditions, the Einstein's GR is
reproduced.

 One of the main
ingredients of the  model\cite{GK1}  consists in the realization
of the idea \cite{Carroll} that the fifth force problem might be
resolved if the theory would possess the approximate global shift
symmetry of the scalar field
\begin{equation}
  \phi\rightarrow\phi +const.\label{phiconst}
\end{equation}
In the  model \cite{GK1},\cite{GK2}, the global shift symmetry
(\ref{phiconst}) is spontaneously broken in such a way that the
effective potential depends on $\phi$ {\it only via}
$M^{4}e^{-2\alpha\phi/M_{p}}$ where $M$ is an integration constant
of the dimensionality of mass that appears as a result of the
spontaneous breakdown of the shift symmetry (\ref{phiconst}). Here
$\alpha >0$ is a parameter of the order of unity and $M_{p}=(8\pi
G)^{-1/2}$. This is a way the model \cite{GK1},\cite{GK2} avoids
the problem with realization of the global shift symmetry
(\ref{phiconst}) in the context of quintessence type models where
the potential is not invariant under the shift of $\phi$. The
model with such features was constructed in the framework of the
Two Measures Field Theory (TMT) \cite{TMT}-\cite{Paradigm}.

In the present paper we show that  the main results  concerning
the decoupling and the restoration of the Einstein's GR in the
model \cite{GK1},\cite{GK2} for fermions (which is rather
complicated), remain also true in a macroscopic description of
matter (which is significantly simpler). This should make more
clear the way of resolution of the fifth force problem in scale
invariant TMT models. Our underlying model involves the coupling
of the dilaton $\phi$ to dust in such a form that Lagrangians are
quite usual, without any exotic term, and the action is  invariant
under scale transformations accompanied by a corresponding shift
(\ref{phiconst}) of the dilaton. After spontaneous symmetry
breaking (SSB),  the effective picture in the Einstein frame
differs in general very much from the Einstein's GR. But if the
local matter density is many orders of magnitude larger then the
vacuum energy density then  Einstein's GR is reproduced, and the
dilaton to matter coupling practically disappears without fine
tuning of the parameters.

\section{Basis of Two Measures Field Theory and
Formulation of the  Scale Invariant Model}
\subsection{Main ideas of the Two Measures Field Theory}

TMT is a generally coordinate invariant theory where {\it all the
difference from the standard field theory in curved space-time
consists only of the following three additional assumptions}:

1. The first assumption is the hypothesis that the effective
action at the energies below the Planck scale has to be of the
form\cite{TMT}-\cite{Paradigm}
\begin{equation}
    S = \int L_{1}\Phi d^{4}x +\int
L_{2}\sqrt{-g}d^{4}x \label{S}
\end{equation}
 including two Lagrangians $ L_{1}$ and $L_{2}$ and two
measures of integration $\sqrt{-g}$ and $\Phi$. One is the usual
measure of integration $\sqrt{-g}$  in the 4-dimensional
space-time manifold equipped with the metric $g_{\mu\nu}$. Another
is the new measure of integration $\Phi$ in the same 4-dimensional
space-time manifold. The measure  $\Phi$ being  a scalar density
and a total derivative may be defined for example\footnote{
Possible nature of the measure fields $\varphi_{a}$ have been
discussed in Ref. \cite{Foundations}.  It is interesting that the
idea of T.D. Lee on the possibility of dynamical
coordinates\cite{TDLee} may be related to the measure fields
$\varphi_{a}$ too. Another possibility consists of the use of a
totally antisymmetric three index field \cite{Foundations}.} by
means of four scalar fields $\varphi_{a}$ ($a=1,2,3,4$)

\begin{equation}
\Phi
=\varepsilon^{\mu\nu\alpha\beta}\varepsilon_{abcd}\partial_{\mu}\varphi_{a}
\partial_{\nu}\varphi_{b}\partial_{\alpha}\varphi_{c}
\partial_{\beta}\varphi_{d}.
\label{Phi}
\end{equation}

To provide parity conservation one can choose for example one of
$\varphi_{a}$'s to be a pseudoscalar.

2. Generically it is allowed that  $ L_{1}$ and $L_{2}$ will be
functions of all matter fields, the dilaton field, the metric, the
connection but not of the "measure fields" $\varphi_{a}$ . In such
a case, i.e. when the measure fields  enter in the theory only via
the measure $\Phi$, the action (\ref{S}) possesses an infinite
dimensional symmetry
$\varphi_{a}\rightarrow\varphi_{a}+f_{a}(L_{1})$, where
$f_{a}(L_{1})$ are arbitrary functions of  $L_{1}$ (see details in
Ref. \cite{DM}). One can hope that this symmetry should prevent
emergence of a measure fields dependence in $ L_{1}$ and $L_{2}$
after quantum effects are taken into account.

3. Important feature of TMT that is responsible for many
interesting and desirable results of the field theory models
studied so far\cite{GK1},\cite{GK2},\cite{TMT}-\cite{Paradigm}
 consists of the assumption that all fields, including
also metric, connection  and the {\it measure fields}
$\varphi_{a}$  are independent dynamical variables. All the
relations between them are results of equations of motion.  In
particular, the independence of the metric and the connection
means that we proceed in the first order formalism and the
relation between connection and metric is not a priori according
to Riemannian geometry.

We want to stress again that except for the listed three
assumptions we do not make any changes as compared with principles
of the standard field theory in curved space-time. In other words,
all the freedom in constructing different models in the framework
of TMT consists of the choice of the concrete matter content and
the Lagrangians $ L_{1}$ and $L_{2}$ that is quite similar to the
standard field theory.

Since $\Phi$ is a total derivative, a shift of $L_{1}$ by a
constant, $L_{1}\rightarrow L_{1}+const$, has no effect on the
equations of motion. Similar shift of $L_{2}$ would lead to the
change of the constant part of the Lagrangian coupled to the
volume element $\sqrt{-g}d^{4}x $. In the standard GR, this
constant term is the cosmological constant. However in TMT the
relation between the constant
 term of $L_{2}$ and the physical cosmological constant is very non
trivial and this makes
possible\cite{DM},\cite{GK1},\cite{GK2},\cite{Paradigm} to resolve
the cosmological constant problem\footnote{Another way to
construct a measure of integration which is a total derivative was
recently studied by Comelli in Ref. \cite{Comelli}. Using a vector
field (instead of four scalar fields $\varphi_a$ used in TMT) and
proceeding in the second order formalism, it was shown in
\cite{Comelli} that it is possible to overcome  the cosmological
constant problem.}.

Varying the measure fields $\varphi_{a}$, we obtain
\begin{equation}
B^{\mu}_{a}\partial_{\mu}L_{1}=0  \quad where \quad
B^{\mu}_{a}=\varepsilon^{\mu\nu\alpha\beta}\varepsilon_{abcd}
\partial_{\nu}\varphi_{b}\partial_{\alpha}\varphi_{c}
\partial_{\beta}\varphi_{d}.\label{varphiB}
\end{equation}
Since $Det (B^{\mu}_{a}) = \frac{4^{-4}}{4!}\Phi^{3}$ it follows
that if $\Phi\neq 0$,
\begin{equation}
 L_{1}=sM^{4} =const \label{varphi}
\end{equation}
where $s=\pm 1$ and $M$ is a constant of integration with the
dimension of mass. In what follows we make the choice $s=1$.

 One should notice
 {\it the very important differences of
TMT from scalar-tensor theories with nonminimal coupling}: \\
 a) In general, the Lagrangian density $L_{1}$ (coupled to the measure
$\Phi$) may contain not only the scalar curvature term (or more
general gravity term) but also all possible matter fields terms.
This means that TMT modifies in general both the gravitational
sector  and the matter sector; \\ b) If the field $\Phi$ were the
fundamental (non composite) one then instead of (\ref{varphi}),
the variation of $\Phi$ would result in the equation $L_{1}=0$ and
therefore the dimensionfull integration constant $M^4$ would not
appear in the theory.

Applying the Palatini formalism in TMT one can show (see for
example \cite{DM})  that in addition to the usual Christoffel
coefficients, the resulting relation between metric and connection
includes also the gradient of the ratio of the two measures
\begin{equation}
\zeta \equiv\frac{\Phi}{\sqrt{-g}} \label{zeta}
\end{equation}
which is a scalar field. This means that with the set of variables
used in the underlying action (\ref{S}) (and in particular with
the metric $g_{\mu\nu}$) the space-time is not Riemannian. The
gravity and matter field equations obtained by means of the first
order formalism contain both $\zeta$ and its gradient. It turns
out that at least at the classical level, the measure fields
$\varphi_{a}$ affect the theory only through the scalar field
$\zeta$.

Variation with respect to the metric yields as usual the
gravitational equations. But in addition, if $L_1$ involves a
 scalar curvature term  then Eq.(\ref{varphi})
 provides us with an additional gravitational type equation, independent
 of the former.
 Taking trace of the gravitational equations and  excluding the scalar
 curvature from these independent equations we obtain a consistency condition
  having the form of a constraint which determines $\zeta (x)$ as a function of
matter fields. It is very important
 that neither Newton constant nor curvature appear in this constraint
which means that the {\it geometrical scalar field} $\zeta (x)$
{\it is determined by other fields configuration}  locally and
straightforward (that is without gravitational interaction).

By an appropriate change of the dynamical variables which includes
a  redefinition of the metric, one can formulate the theory in a
Riemannian space-time. The corresponding  frame we call "the
Einstein frame". The big advantage of TMT is that in a very wide
class of models, {\it the gravity and all matter fields equations
of motion take canonical GR form in the Einstein frame}.
 All the novelty of TMT in the Einstein frame as compared
with the standard GR is revealed only
 in an unusual structure of the scalar fields
effective potential, masses of particles  and their interactions
with scalar fields as well as in the unusual structure of matter
contributions to the energy-momentum tensor: all these quantities
appear to be $\zeta$ dependent. This is why the scalar field
$\zeta (x)$ determined by the constraint as a function of matter
fields, has a key role in dynamics of TMT models. Note that if we
were to assume that for some reasons the gravity effects are
negligible and choose to work in the Minkowski space-time from the
very beginning, then we would lose the constraint, and the result
would be very much different from the one obtained according to
the prescriptions of TMT with
 taking the flat space-time limit at the end. This means that the
 gravity in TMT plays the more essential role than in the usual
 (i.e. only with the measure of integration $\sqrt{-g}$)
 field theory in curved space-time.

\subsection{Scale invariant model}

In the original frame (where the metric is $g_{\mu\nu}$), a matter
content of our TMT model represented in the form of the action
(\ref{S}), is a dust and a scalar field (dilaton). The dilaton
$\phi$ allows to realize a spontaneously broken global scale
invariance\cite{G1},\cite{G2},\cite{GK1},\cite{GK2} and together
with this it can govern the evolution of the universe on different
stages: in the early universe $\phi$ plays the role of inflaton
and in the late time universe it is transformed into a part of the
dark energy (for details see Refs.
\cite{GK1},\cite{GK2},\cite{Paradigm}). We postulate that the
theory is invariant under the global scale transformations:
\begin{equation}
    g_{\mu\nu}\rightarrow e^{\theta }g_{\mu\nu}, \quad
\Gamma^{\mu}_{\alpha\beta}\rightarrow \Gamma^{\mu}_{\alpha\beta},
\quad \phi\rightarrow \phi-\frac{M_{p}}{\alpha}\theta, \quad
\varphi_{a}\rightarrow l_{ab}\varphi_{b} \label{st}
\end{equation}
where $\det(l_{ab})=e^{2\theta}$ and $\theta =const$. Keeping the
general structure (\ref{S}), it is convenient to represent the
action in the following form:
\begin{eqnarray}
S&=&S_g+S_{\phi}+S_{m} \label{SgSphiSm}
\label{totaction}\\
 S_g&=&-\frac{1}{\kappa}\int (\Phi +b_{g}\sqrt{-g})R(\Gamma,g) e^{\alpha\phi
 /M_{p}}d^{4}x \,;
\nonumber\\
S_{\phi}&=&\int e^{\alpha\phi/M_{p}}\left[(\Phi
+b_{\phi}\sqrt{-g})\frac{1}{2}g^{\mu\nu}\phi_{,\mu}\phi_{,\nu}-
 \left(\Phi V_{1}
+\sqrt{-g}V_{2}\right)e^{\alpha\phi /M_{p}}\right]d^{4}x \,;
\nonumber\\
S_{m}&=&\int (\Phi +b_{m}\sqrt{-g})L_m d^{4}x \, , \nonumber
\nonumber
\end{eqnarray}
where
\begin{eqnarray}
R(\Gamma,g)=g^{\mu\nu}\left( \Gamma^{\lambda}_{\mu\nu
,\lambda}-\Gamma^{\lambda}_{\mu\lambda ,\nu}+
\Gamma^{\lambda}_{\alpha\lambda}\Gamma^{\alpha}_{\mu\nu}-
\Gamma^{\lambda}_{\alpha\nu}\Gamma^{\alpha}_{\mu\lambda}\right)
\nonumber
\end{eqnarray}
and the Lagrangian for the matter, as collection of particles,
which provides the scale invariance of $S_m$ reads
\begin{equation}
L_m=-m\sum_{i}\int e^{\frac{1}{2}\alpha\phi/M_{p}}
\sqrt{g_{\alpha\beta}\frac{dx_i^{\alpha}}{d\lambda}\frac{dx_i^{\beta}}{d\lambda}}\,
\frac{\delta^4(x-x_i(\lambda))}{\sqrt{-g}}d\lambda \label{Lm}
\end{equation}
where $\lambda$ is an arbitrary parameter. For simplicity we
consider the collection of the particles with the same mass
parameter $m$. We assume in addition that $x_i(\lambda)$ do not
participate in the scale transformations (\ref{st}).

In the action (\ref{totaction}) there are two types of the
gravitational terms and
 of the "kinetic-like terms"  which
respect the scale invariance : the terms of the one type coupled
to the
 measure $\Phi$ and those of the other type
coupled to the measure $\sqrt{-g}$. Using the freedom in
normalization of the measure fields $\varphi_{a}$   we set the
coupling constant of the scalar curvature to the measure $\Phi$ to
be  $-\frac{1}{\kappa}$. Normalizing all the fields such that
their couplings to the measure $\Phi$ have no additional factors,
we are not able in general to provide the same in terms describing
the appropriate couplings to the measure $\sqrt{-g}$. This fact
explains the need to introduce the dimensionless real parameters
$b_g$, $b_{\phi}$ and $b_m$. We will only assume that  they are
positive, have the same or very close orders of magnitude
\begin{equation}
b_g\sim b_{\phi}\sim b_m \label{sim-bg-bm-bphi}
\end{equation}
and besides  $b_m>b_g$. The real positive parameter $\alpha$ is
assumed to be of the order of unity. As usual  $\kappa =16\pi G $
and we use $M_p=(8\pi G)^{-1/2}$.

One should  also point out the possibility of introducing two
different pre-potentials which are exponential functions of the
dilaton $\phi$ coupled to the measures $\Phi$ and $\sqrt{-g}$ with
factors $V_{1}$ and $V_{2}$. Such $\phi$-dependence provides the
scale symmetry (\ref{st}). We will see below how the dilaton
effective potential is generated as the result of SSB of the scale
invariance and the transformation to the  Einstein frame.

According to the general prescriptions of TMT, we have to start
from studying the self-consistent system of gravity (metric
$g_{\mu\nu}$ and connection $\Gamma^{\mu}_{\alpha\beta}$), the
measure $\Phi$ degrees of freedom $\varphi_a$, the dilaton field
$\phi$ and the matter particles coordinates
$x^{\alpha}_i(\lambda)$, proceeding in the first order formalism.

For the purpose of this paper we restrict ourselves to a zero
temperature gas of particles, i.e. we will assume that
$d\vec{x}_i/d\lambda \equiv 0$  for all particles. It is
convenient to proceed in the frame where $g_{0l}=0$, \, $l=1,2,3$.
Then the particle density is defined by
\begin{equation}
n(\vec{x})=\sum_{i}\frac{1}{\sqrt{-g_{(3)}}}\delta^{(3)}(\vec{x}-\vec{x}_i(\lambda))
\label{n(x)}
\end{equation}
where $g_{(3)}=\det(g_{kl})$ and
\begin{equation}
S_{m}=-m\int d^{4}x(\Phi
+b_{m}\sqrt{-g})\,n(\vec{x})\,e^{\frac{1}{2}\alpha\phi/M_{p}}
\label{S-n(x)}
\end{equation}

Following the procedure described in the previous subsection we
have to write down all equations of motion, find the consistency
condition (the constraint which determines $\zeta$-field as a
function of other fields and matter) and make a transformation to
the Einstein frame. We will skip most of the intermediate results
and in the next subsection  present the resulting equations in the
Einstein frame. Nevertheless  two exclusions we have to make here.

The first one concerns the important effect observable when
varying $S_m$ with respect to $g^{\mu\nu}$:
\begin{equation}
\frac{\delta S_{m}}{\delta g^{00}}=\frac{b_m}{2}\,\sqrt{-g} \, m
\, n(\vec{x}) \, e^{\frac{1}{2}\alpha\phi/M_{p}}\, g_{00},
\label{00var}
\end{equation}
\begin{equation}
\frac{\delta S_{m}}{\delta g^{kl}}=-\frac{1}{2}\,\Phi\, m \,
n(\vec{x}) \, e^{\frac{1}{2}\alpha\phi/M_{p}}\, g_{kl}.
\label{klvar}
\end{equation}
The latter equation shows that due to the measure $\Phi$, {\em the
zero temperature gas generically possesses a pressure}. As we will
see this pressure  disappears automatically together with the
fifth force as the matter energy density is many orders of
magnitude larger then the dark energy density, which is evidently
true in all physical phenomena tested experimentally.

The second one  is the notion concerning the role of Eq.
(\ref{varphi}) resulting from variation of the measure fields
$\varphi_a$. With the action (\ref{SgSphiSm}), where the
Lagrangian $L_1$ is the sum of terms coupled to the measure
$\Phi$, Eq. (\ref{varphi}) describes a {\em spontaneous breakdown
of the global scale symmetry} (\ref{st}).

\section{Equations of motion in the Einstein frame.}

It turns out that when working with the new metric ($\phi$
 remains the same)
\begin{equation}
\tilde{g}_{\mu\nu}=e^{\alpha\phi/M_{p}}(\zeta +b_{g})g_{\mu\nu},
\label{ct}
\end{equation}
which we call the Einstein frame,
 the connection  becomes Riemannian.
  Since
$\tilde{g}_{\mu\nu}$ is invariant under the scale transformations
(\ref{st}), spontaneous breaking of the scale symmetry (by means
of Eq.(\ref{varphi}))  is reduced in the Einstein frame to the
{\it spontaneous breakdown of the shift symmetry}
(\ref{phiconst}). Notice that the Goldstone theorem generically is
not applicable in this kind of models\cite{G2}.

 The transformation
(\ref{ct}) causes the transformation of the particle density
\begin{equation}
\tilde{n}(\vec{x})=(\zeta +b_g)^{-3/2}\,
e^{-\frac{3}{2}\alpha\phi/M_{p}}\, n(\vec{x}) \label{ntilde}
\end{equation}

 After the change of
variables  to the Einstein frame (\ref{ct}) and some simple
algebra, the gravitational equations take the standard GR form
\begin{equation}
G_{\mu\nu}(\tilde{g}_{\alpha\beta})=\frac{\kappa}{2}T_{\mu\nu}^{eff}
 \label{gef}
\end{equation}
where  $G_{\mu\nu}(\tilde{g}_{\alpha\beta})$ is the Einstein
tensor in the Riemannian space-time with the metric
$\tilde{g}_{\mu\nu}$. The components of the effective
energy-momentum tensor are as follows
\begin{eqnarray}
T_{00}^{eff}&=&\frac{\zeta +b_{\phi}}{\zeta
+b_{g}} \left(\dot{\phi}^2- \tilde{g}_{00}X\right) \label{T00}\\
&+&\tilde{g}_{00}\left[V_{eff}(\phi;\zeta,M)-\frac{\delta\cdot
b_g}{\zeta +b_{g}}X+\frac{3\zeta +b_m +2b_g}{2\sqrt{\zeta
+b_{g}}}\, m\, \tilde{n}\right] \nonumber
\end{eqnarray}
\begin{eqnarray}
T_{ij}^{eff}&=&\frac{\zeta +b_{\phi}}{\zeta +b_{g}}
\left(\phi_{,k}\phi_{,l}-\tilde{g}_{kl}X\right)
\label{Tkl}\\
&+&\tilde{g}_{kl}\left[V_{eff}(\phi;\zeta,M)-\frac{\delta\cdot
b_g}{\zeta +b_{g}}X+\frac{\zeta -b_m +2b_g}{2\sqrt{\zeta
+b_{g}}}\, m\, \tilde{n}\right]\nonumber
\end{eqnarray}
Here the following notations have been used:
\begin{equation}
X\equiv\frac{1}{2}\tilde{g}^{\alpha\beta}\phi_{,\alpha}\phi_{,\beta}
\qquad and \qquad \delta =\frac{b_{g}-b_{\phi}}{b_{g}}
\label{delta}
\end{equation}
and the function $V_{eff}(\phi ;\zeta)$ is defined by
\begin{equation}
V_{eff}(\phi ;\zeta)=
\frac{b_{g}\left[M^{4}e^{-2\alpha\phi/M_{p}}+V_{1}\right]
-V_{2}}{(\zeta +b_{g})^{2}} \label{Veff1}
\end{equation}

The dilaton $\phi$ field equation in the Einstein frame is as
follows
\begin{eqnarray}
&&\frac{1}{\sqrt{-\tilde{g}}}\partial_{\mu}\left[\frac{\zeta
+b_{\phi}}{\zeta
+b_{g}}\sqrt{-\tilde{g}}\tilde{g}^{\mu\nu}\partial_{\nu}\phi\right]\nonumber\\
&&-\frac{\alpha}{M_{p}}\,\frac{(\zeta
+b_{g})M^{4}e^{-2\alpha\phi/M_{p}}-(\zeta -b_{g})V_{1}
-2V_{2}-\delta b_{g}(\zeta
+b_{g})X}{(\zeta +b_{g})^{2}}\nonumber\\
 &&=\frac{\alpha}{M_{p}}\,\frac{\zeta
-b_m +2b_g}{2\sqrt{\zeta +b_{g}}}\, m\,\tilde{n}
 \label{phief}
\end{eqnarray}

In the above equations, the scalar field $\zeta$  is determined as
a function $\zeta(\phi,X,\tilde{n})$ by means of the following
constraint (origin of which has been discussed in Sec.2.1):
\begin{eqnarray}
&&\frac{(b_{g}-\zeta)\left(M^{4}e^{-2\alpha\phi/M_{p}}+V_{1}\right)-2V_{2}}{(\zeta
+b_g)^2} -\frac{\delta\cdot b_{g}X}{\zeta +b_g}=
\frac{\zeta -b_m +2b_g}{2\sqrt{\zeta +b_{g}}}\, m\, \tilde{n}\nonumber\\
\label{constraint2}
\end{eqnarray}

Applying the constraint (\ref{constraint2}) to Eq.(\ref{phief})
one can reduce the latter to the form
\begin{equation}
\frac{1}{\sqrt{-\tilde{g}}}\partial_{\mu}\left[\frac{\zeta
+b_{\phi}}{\zeta
+b_{g}}\sqrt{-\tilde{g}}\tilde{g}^{\mu\nu}\partial_{\nu}\phi\right]-\frac{2\alpha\zeta}{(\zeta
+b_{g})^{2}M_{p}} M^{4}e^{-2\alpha\phi/M_{p}} =0,
\label{phi-after-con}
\end{equation}
where $\zeta$  is a solution of the constraint
(\ref{constraint2}).

 One should
point out two very important features of the model. First, the
$\phi$ dependence in all the equations of motion (including the
constraint) emerges only in the form $M^{4}e^{-2\alpha\phi/M_{p}}$
where $M$ is the integration constant, i.e. due to the spontaneous
breakdown of the scale symmetry (\ref{st}) (or the shift symmetry
(\ref{phiconst}) in the Einstein frame). Second, the constraint
(\ref{constraint2}) is the fifth degree algebraic equation with
respect to $\sqrt{\zeta +b_g}$ and therefore generically $\zeta$
is a complicated function of  $\phi$, $X$ and $\tilde{n}$. Hence
generically each of $\zeta$ dependent terms in Eqs.
(\ref{T00})-(\ref{phief}) and (\ref{phi-after-con}) describe very
nontrivial coupling of the dilaton to the matter.

\section{Dark Energy in the Absence of Matter}

It is worth to start the investigation of the features of our
model from the simplest case when the particle density of the dust
is zero: $\tilde{n}(x)\equiv 0$. Then the dilaton $\phi$ is
 the only matter which in the early universe plays the
role of the inflaton while in the late universe it is the dark
energy. The appropriate model in the context of cosmological
solutions has been studied in detail in Ref. \cite{Paradigm}. Here
we present only some of the equations we will need for the
purposes of this paper and a list of the main results.

In the absence of the matter particles, the scalar $\zeta
=\zeta(\phi,X)$ can be easily
  found  from the constraint (\ref{constraint2}):
\begin{equation}
\zeta^{(\tilde{n}=0)}=b_g-2\frac{V_2+\delta\cdot
b_g^2X}{M^{4}e^{-2\alpha\phi/M_{p}}+V_1+\delta\cdot b_gX}
\label{zeta-DE}
\end{equation}
In the spatially homogeneous case $X\geq0$. Then the effective
energy-momentum tensor can be represented in a form of that of  a
perfect fluid
\begin{equation}
T_{\mu\nu}^{eff}=(\rho +p)u_{\mu}u_{\nu}-p\tilde{g}_{\mu\nu},
\qquad where \qquad
u_{\mu}=\frac{\phi_{,\mu}}{(2X)^{1/2}}\label{Tmnfluid}
\end{equation}
with the following energy and pressure densities
 obtained after inserting
(\ref{zeta-DE}) into the components of the energy-momentum tensor
(\ref{T00}), (\ref{Tkl}) where now $\tilde{n}(x)\equiv 0$
\begin{eqnarray}
&&\rho(\phi,X;M)\equiv \rho^{(\tilde{n}=0)} \label{rho1}\\
&&=X+\frac{(M^{4}e^{-2\alpha\phi/M_{p}}+V_{1})^{2}- 2\delta
b_{g}(M^{4}e^{-2\alpha\phi/M_{p}}+V_{1})X -3\delta^{2}
b_{g}^{2}X^2}{4[b_{g}(M^{4}e^{-2\alpha\phi/M_{p}}+V_{1})-V_{2}]},
\nonumber
\end{eqnarray}
\begin{equation}
p(\phi,X;M)\equiv p^{(\tilde{n}=0)} =X-
\frac{\left(M^{4}e^{-2\alpha\phi/M_{p}}+V_{1}+ \delta
b_{g}X\right)^2}
{4[b_{g}(M^{4}e^{-2\alpha\phi/M_{p}}+V_{1})-V_{2}]}. \label{p1}
\end{equation}
Substitution of (\ref{zeta-DE}) into Eq. (\ref{phi-after-con})
yields the $\phi$-equation with very interesting dynamics. The
appearance of the nonlinear $X$ dependence in spite of the absence
of such nonlinearity in the underlying action means that our model
represents an  explicit example of $k$-essence\cite{k-essence}
resulting from first principles. The effective $k$-essence action
has the form
\begin{equation}
S_{eff}=\int\sqrt{-\tilde{g}}d^{4}x\left[-\frac{1}{\kappa}R(\tilde{g})
+p\left(\phi,X;M\right)\right] \label{k-eff},
\end{equation}
where $p(\phi,X;M)$ is given by Eq.(\ref{p1}).

In the context of spatially flat FRW cosmology, in the absence of
the matter particles (i.e $\tilde{n}(x)\equiv 0$), the TMT model
under consideration\cite{Paradigm} exhibits a number of
interesting
 outputs depending of the choice of regions in the parameter space (but
without fine tuning): \\
a) {\it Absence of initial singularity of the curvature while its
time derivative is singular}. This is a sort of "sudden"
singularities studied by Barrow on purely kinematic grounds\cite{Barrow}. \\
b) Power law inflation in the subsequent stage of evolution.
Depending on the region in the parameter space the inflation ends
with a {\it graceful exit} either into the state with zero
cosmological constant (CC) or into the state driven by both a
small CC and the field $\phi$ with a quintessence-like
potential.\\
c) Possibility of {\it resolution of the old CC problem}. From the
point of view of TMT, it becomes clear why the old CC problem
cannot be solved (without fine tuning) in conventional field
theories.\\
 d) TMT enables two ways for achieving small CC without
fine tuning of dimensionful parameters: either by a {\it seesaw}
type mechanism or due to
 {\it a correspondence principle} between TMT and
conventional field theories (i.e theories with only the measure of
integration $\sqrt{-g}$  in the action).\\
  e) There  is a wide range of the
parameters where the dynamics of the scalar field $\phi$, playing
the role of the dark energy in the late universe, allows crossing
the phantom divide, i.e. the equation-of-state $w=p/\rho$ may be
$w<-1$ and $w$  asymptotically (as $t\rightarrow\infty$)
approaches $-1$ from below. One can show that in the original
frame used in the underlying action (\ref{totaction}), this regime
corresponds to the negative sign of the measure of integration
$\Phi +b_{\phi}\sqrt{-g}$ of the dilaton $\phi$ kinetic
term\footnote{Note that by the definition (\ref{Phi}) the measure
$\Phi$ is not positive definite. In the Measure Theory the
non-positive definite measure is known as "Signed Measure", see
for example Ref. \cite{MeasureTheory}.}. This dynamical effect
which emerges here instead of putting the wrong sign kinetic term
by hand in the phantom model\cite{Phantom}, will be discussed in
detail in another paper.

Taking into account that in the late time universe  the
$X$-contribution to $\rho^{(\tilde{n}=0)}$ approaches zero, one
can see that the dark energy density is positive for any $\phi$
provided
\begin{equation}
 b_{g}V_{1}\geq V_{2} \label{bV1>V2}
\end{equation}
Then it follows from (\ref{zeta-DE}) that
\begin{equation}
 |\zeta^{(\tilde{n}=0)}|\sim b_g. \label{zeta-sim-bg}
\end{equation}
 This will be useful in the next
section.

\section{Normal Conditions:
Reproducing Einstein's GR and Absence of the Fifth Force Problem}

One should now pay attention to the interesting result that  the
explicit $\tilde{n}$ dependence involving {\bf the  same form of
$\zeta$ dependence}
\begin{equation}
 \frac{\zeta -b_m +2b_g}{2\sqrt{\zeta +b_{g}}}\, m\,
\tilde{n} \label{universality}
\end{equation}
 appears simultaneously\footnote{Note that analogous result has been observed
earlier in the model\cite{GK1},\cite{GK2} where fermionic matter
has been studied instead of the macroscopic (dust) matter in the
present model.} in the dust contribution to the pressure (through
the last term in Eq. (\ref{Tkl})), in the  effective dilaton to
dust coupling (in the r.h.s. of Eq. (\ref{phief})) and in the
r.h.s. of the constraint (\ref{constraint2}).

Let us analyze consequences of this wonderful coincidence in the
case when the matter energy density (modeled by
 dust) is much larger than the dilaton contribution to the dark
 energy density in the space region occupied by this matter. Evidently
this is the condition under which all tests of Einstein's GR,
including the question of the fifth force, are fulfilled.
Therefore if this condition is satisfied we will say that the
matter is in {\bf normal conditions}. The existence of the fifth
force turns into a problem just in normal conditions. The opposite
situation may be realized (see Refs. \cite{GK1},\cite{GK2}) if the
matter is diluted up to a magnitude of the macroscopic energy
density comparable with the dilaton contribution to the dark
energy density. In this case we
 say that the matter is in
 the  state of cosmo-low energy physics ({\bf CLEP}). It is
 evident that the fifth force acting on the matter in the CLEP
 state cannot be detected now and in the near future, and therefore
 does not appear to be a problem. But effects of the CLEP may be
 important in cosmology, see Ref. \cite{GK2}.

  The last terms in eqs. (\ref{T00}) and (\ref{Tkl}),
 being the matter contributions to the energy density ($\rho_m$) and the
 pressure ($-p_m$) respectively, generally speaking have the same
 order of magnitude. But if the dust is in the normal conditions
 there is a possibility to provide the desirable feature of the dust in GR: it
 must be pressureless. This is realized provided that in normal
 conditions (n.c.) the following equality holds with extremely
 high accuracy:
\begin{equation}
 \zeta^{(n.c.)}\approx b_m-2b_g
\label{decoupling-cond}
\end{equation}
Remind that we have assumed  $b_m >b_g$. Then $\zeta^{(n.c.)}+b_g
>0$, and the transformation (\ref{ct}) and the subsequent
equations in the Einstein frame are well defined.
 Inserting
(\ref{decoupling-cond}) in the last term of Eq. (\ref{T00}) we
obtain the effective dust energy density in normal conditions
\begin{equation}
 \rho_m^{(n.c.)}=2\sqrt{b_m-b_g} \, m\tilde{n}
\label{rho-m-n.c.}
\end{equation}
Substitution of (\ref{decoupling-cond}) into the rest of the terms
of the components of the energy-momentum tensor (\ref{T00}) and
(\ref{Tkl}) gives the dilaton contribution to the energy density
and pressure of the dark energy which have the orders of magnitude
close to those in the absence of matter case, Eqs. (\ref{rho1})
and (\ref{p1}). The latter statement may be easily checked by
using Eqs.
(\ref{zeta-DE}),(\ref{zeta-sim-bg}),(\ref{decoupling-cond}) and
(\ref{sim-bg-bm-bphi}).

Note that Eq. (\ref{decoupling-cond}) is not just a choice to
provide zero dust contribution to the pressure. In fact it is the
result of analyzing the equations of motion together with the
constraint (\ref{constraint2}). In the Appendix we present the
detailed analysis yielding this result. But in this section we
have started from the use of this result in order to make the
physical meaning more distinct.

Taking into account our assumption (\ref{sim-bg-bm-bphi}) and Eq.
(\ref{zeta-sim-bg}) we infer that $\zeta^{(n.c.)}$ and
$\zeta^{(\tilde{n}=0)}$ (in the absence of matter case, Eq.
(\ref{zeta-DE})) have close orders of magnitudes. Then it is easy
to see (making use the inequality (\ref{bV1>V2})) that the l.h.s.
of the constraint (\ref{constraint2}), as $\zeta =\zeta^{(n.c.)}$,
has the order of magnitude close to that of the dark energy
density $\rho^{(\tilde{n}=0)}$ in the absence of matter case
discussed in Sec. 4. Thus in the case under consideration, {\em
the constraint (\ref{constraint2}) describes a balance between the
pressure of the dust in normal conditions on the one hand and the
vacuum energy density on the other hand}. This balance is realized
due to the condition (\ref{decoupling-cond}).

 Besides reproducing Einstein equations when
the scalar field and dust (in normal conditions) are sources of
the gravity, {\em the condition (\ref{decoupling-cond})
automatically provides a practical disappearance of the effective
dilaton to matter coupling}. Indeed, inserting
(\ref{decoupling-cond}) into the $\phi$-equation written in the
form (\ref{phi-after-con}) and into $V_{eff}(\phi ;\zeta)$, Eq.
(\ref{Veff1}), one can immediately see that only the force of the
strength of the dark energy selfinteraction is present in this
case. Note that this force is a total force involving both the
selfinteraction of the dilaton and its interaction with dust in
normal conditions. Furthermore, in this way one can see explicitly
that due to the factor $M^{4}e^{-2\alpha\phi/M_{p}}$, this total
force may obtain an additional, exponential  dumping since in the
cosmological context shortly discussed in Sec. 4 (see details in
Ref. (\cite{Paradigm})) a scenario, where in the late time
universe $\phi\gg M_p$, seems to be most appealing.

Another way to see the absence of the fifth force problem in the
normal conditions is to look at the $\phi$-equation in the form
(\ref{phief}) and estimate the Yukawa type coupling constant in
the r.h.s. of this equation. In fact, using the constraint
(\ref{constraint2}) and representing the particle density in the
form $\tilde{n}\approx N/\upsilon$ where $N$ is the number of
particles in a volume $\upsilon$, one can make the following
estimation for the effective dilaton to matter coupling "constant"
$f$ defined by the Yukawa type interaction term $f\tilde{n}\phi$
(if we were to invent an effective action whose variation with
respect to $\phi$ would result in Eq. (\ref{phief})):
\begin{equation}
f \equiv\alpha\frac{m}{M_{p}}\,\frac{\zeta -b_m
+2b_g}{2\sqrt{\zeta +b_{g}}}\approx
\alpha\frac{m}{M_{p}}\,\frac{\zeta -b_m
+2b_g}{2\sqrt{b_m-b_{g}}}\sim
\frac{\alpha}{M_{p}}\,\frac{\rho_{vac}}{\tilde{n}} \approx
\alpha\frac{\rho_{vac}\upsilon}{NM_{p}} \label{Archimed}
\end{equation}
 Thus we conclude that {\it the
effective coupling "constant" of the dilaton to matter in the
normal conditions is of the order of the ratio of  the "mass of
the vacuum" in the volume occupied by the matter to   the Planck
mass taken $N$ times}. In some sense this result resembles the
{\it Archimedes law}. At the same time Eq. (\ref{Archimed}) gives
us an estimation of the exactness of the condition
(\ref{decoupling-cond}).

\section{Discussion and Conclusion}

In the  present paper, the idea to construct a model with
spontaneously broken dilatation invariance where the dilaton
dependence in all equations of motion results only from the SSB of
the shift symmetry (\ref{phiconst}), is implemented from first
principles in the framework of TMT.

Although the dust  model studied in this paper is a very crude
model of matter, it is quite sufficient for studying the fifth
force problem. In fact, all experiments which search for the fifth
force deal with macroscopic bodies which, in the zeroth order
approximation, can be regarded as collections of noninteracting,
point-like motionless particles with very high particle number
density $\tilde{n}(x)$.

Generically the model studied in the present paper is different
from  Einstein's GR. For example  it allows the long range scalar
force and a non-zero pressure of the cold dust. However the
magnitude of the particle number density turns out to be the very
important factor influencing the strength of the dilaton to matter
coupling. This happens due to the constraint (\ref{constraint2})
which is nothing but the consistency condition of the equations of
motion. The analysis of the constraint presented in the Appendix
shows that generically it describes a balance between the matter
density and dark energy density. It turns out that in the case of
a macroscopic body, that is in normal conditions, the constraint
allows this balance only in such a way that the dilaton
practically decouples from the matter and  Einstein's GR is
restored automatically. Thus our model not only explains why all
attempts to discover a scalar force correction to Newtonian
gravity were unsuccessful so far but also predicts that in the
near future there is no chance to detect such corrections  in the
astronomical measurements as well as in the specially designed
fifth force experiments on intermediate, short (like millimeter)
and even ultrashort (a few nanometer) ranges. This prediction is
alternative to predictions of other known models.

Formally one can consider the case of a very diluted matter when
the matter energy density is of the order of magnitude comparable
with the dark energy density, which is the case opposite to the
normal conditions. Only in this case the balance dictated by the
constraint implies the existence of a non small dilaton coupling
to matter, as well as a possibility of other distinctions from
Einstein's GR. However these effects cannot be detected in fifth
force experiments now and in the near future. One should also note
here that in the framework of the present model based on the
consideration of point particles, the low density limit, strictly
speaking, cannot be satisfactory defined. An example of the
appropriate low density limit (CLEP state)  was realized using a
field theory model in Ref. \cite{GK2} while conclusions for matter
in the normal conditions were very similar to results of the
present paper.

Possible cosmological and astrophysical effects when the normal
conditions are not satisfied may be very interesting. In
particular, taking into account that all dark matter known in the
present universe has the macroscopic energy density many orders of
magnitude smaller than the energy density of visible macroscopic
bodies, we hope that the nature of the dark matter can be
understood as a state opposite to the normal conditions studied in
the present paper.

\section{Appendix. $\zeta(x)$ when the matter is in normal conditions}

As we mentioned in Sec. 3, solutions $\zeta
=\zeta(\phi,X,\tilde{n})$ of the constraint (\ref{constraint2})
are generically very complicated functions. Nevertheless let us
imagine that we solve the constraint, substitute $\zeta
=\zeta(\phi,X,\tilde{n})$ into eqs. (\ref{T00}), (\ref{Tkl}),
(\ref{phief}) and solve them with certain boundary or/and initial
conditions. Inserting the obtained solutions for $\phi(x)$ and
$X(x)$ back into $\zeta =\zeta(\phi,X,\tilde{n})$ we will obtain a
space-time dependence of the scalar field $\zeta =\zeta(x)$.

Let us analyze possible regimes for the $\zeta(x)$ having in mind
its possible numerical values.  As we have seen at the end of Sec.
4, in the vacuum $|\zeta^{(\tilde{n}=0)}|\sim b_g$. One can start
asking the following question: what is the effect of inserting
dust (into a vacuum) on the magnitude of $\zeta(x)$ in comparison
with $\zeta^{(\tilde{n}=0)}$? One can think of three possible
regimes: $|\zeta(x)|$ may become significantly larger  than $b_g$,
may keep the same order of magnitude $|\zeta(x)|\sim b_g$ as it
was in the vacuum and may become significantly less than $b_g$.
Consider each of these possibilities.
\begin{enumerate}

\item $\zeta(x)\gg b_g$ \quad Let us start from the notion that
if formally $\zeta\to\infty$ then for any particle density
$\tilde{n}\neq 0$, the l.h.s. of the constraint
(\ref{constraint2}) approaches zero while the r.h.s. approaches
infinity. Therefore a regime where
$\zeta\to\infty$ is impossible.\\
     Consider now the case $\zeta(x)\gg b_g$ with finite
     $\zeta$. We start from estimations of the order of
     magnitude of two terms of $V_{eff}(\phi;\zeta)$, Eq. (\ref{Veff1}), in the vacuum,
     i.e. $V_{eff}(\phi;\zeta)|_{\zeta=\zeta^{(\tilde{n}=0)}}$.
     Using Eq. (\ref{zeta-sim-bg}) we have
     \begin{equation}
\left(\frac{b_g\left[M^{4}e^{-2\alpha\phi/M_{p}}+V_{1}\right]}{(\zeta
+b_g)^{2}}\right)_{vac}\sim
\frac{M^{4}e^{-2\alpha\phi/M_{p}}+V_{1}}{b_g}\label{Estim-1}
\end{equation}
and
\begin{equation}
\left(\frac{|V_{2}|}{(\zeta +b_g)^{2}}\right)_{vac}\sim
\frac{|V_{2}|}{b_g^2} \label{Estim-2}
\end{equation}
In the presence of dust, in the regime $\zeta(x)\gg b_g$ we have
respectively:
\begin{eqnarray}
\left(\frac{b_g\left[M^{4}e^{-2\alpha\phi/M_{p}}+V_{1}\right]}{(\zeta
+b_g)^{2}}\right)_{\tilde{n}\neq 0}\ll
\frac{M^{4}e^{-2\alpha\phi/M_{p}}+V_{1}}{\zeta} \ll
\frac{M^{4}e^{-2\alpha\phi/M_{p}}+V_{1}}{b_g}\label{Estim-3}
\end{eqnarray}
and
\begin{equation}
\left(\frac{|V_{2}|}{(\zeta +b_g)^{2}}\right)_{\tilde{n}\neq
0}\sim \frac{|V_{2}|}{\zeta^2}\ll \frac{|V_{2}|}{b_g^2}
\label{Estim-4}
\end{equation}
Therefore generically
\begin{equation}
V_{eff}(\phi;\zeta)|_{\tilde{n}\neq 0}\ll
V_{eff}(\phi;\zeta)|_{vac}\label{Estim-5}
\end{equation}
where we have ignored possible different values of $\phi$ in the
vacuum and inside the matter\footnote{If $V_1>0$ then  in the late
universe $\phi\gg M_p$, $M^{4}e^{-2\alpha\phi/M_{p}}\ll V_{1}$ and
the universe is driven\cite{Paradigm} mainly by the cosmological
constant $\Lambda =V_{1}^{2}/[4(b_{g}V_{1}-V_{2})]$.}. Further,
proceeding in the same manner with the constraint
(\ref{constraint2}) and using the above estimations it is easily
to see that in the regime $\zeta(x)\gg b_g$, the absolute value of
the l.h.s. of the constraint (\ref{constraint2}) is much less than
the vacuum energy density. But the r.h.s. of the constraint
(\ref{constraint2}) is of the order of the dust contribution to
the energy density (see the last term of Eq. (\ref{T00}) in the
regime $\zeta(x)\gg b_g$). Therefore in normal conditions (large
$\tilde{n}$) the constraint (\ref{constraint2}) does not allow the
regime $\zeta(x)\gg b_g$.

\item $|\zeta(x)|\sim b_g$ \quad In this case the l.h.s. of the
constraint (\ref{constraint2}) has the order of the vacuum energy
density. Let us start from the assumption that $\zeta(x)$, being
$|\zeta(x)|\sim b_g$, is different from the value $\zeta
=b_m-2b_g$. Then the r.h.s. of the constraint (\ref{constraint2}),
being equal to the dust contribution to the pressure (the last
term of Eq. (\ref{Tkl})), has also the order of magnitude of the
dust contribution to the energy density (the last term of Eq.
(\ref{T00})). Therefore in normal conditions (large $\tilde{n}$)
the constraint (\ref{constraint2}) cannot be satisfied if the
value $\zeta$ is far from $b_m-2b_g$. The only way to satisfy the
constraint (\ref{constraint2}) in the regime $|\zeta(x)|\sim b_g$
when the dust is in normal conditions is the equality
(\ref{decoupling-cond}). Consequences of this condition are
discuused in Sec. 5.

\item $|\zeta(x)|\ll b_g$ \quad In this case the l.h.s. of the constraint
 (\ref{constraint2}) has again the order of the vacuum energy
density.  But the r.h.s. of the constraint (\ref{constraint2})
 has generically the same order of magnitude as the dust contribution to the energy
 density (the last term in Eq. (\ref{T00})). Therefore in normal conditions,
 the constraint allows the balance
 (in order of magnitude) between the dark energy density (in the l.h.s. of the constraint)
 and the r.h.s. of the constraint
 provided a tuning of the parameters $b_m\approx
 2b_g$. Thus the regime $|\zeta(x)|\ll b_g$ is a particular case
 of the solution (\ref{decoupling-cond}) if the relation between the parameters $b_g$
 and $b_m$ is about $b_m\approx
 2b_g$.
\end{enumerate}

\end{document}